\documentclass[aps,prd,twocolumn,groupedaddress,amsfonts,amssymb,amsmath,showpacs,showkeys]{revtex4-1}

\usepackage{mathtools}

\usepackage{graphicx}

\usepackage{psfrag}

\newcommand{\aap}{Astronomy and Astrophysics}

\newcommand{\mnras}{MNRAS}

\begin{document}

\title{Astrophysical constraints and insights on extended relativistic gravity.}

\author{S. Mendoza$^{1}$}
\email[Email address: ]{sergio@astro.unam.mx}
\author{Gonzalo J. Olmo$^{2}$}
\email[Email address: ]{gonzalo.olmo@csic.es}
\affiliation{$^1$Instituto de Astronom\'{\i}a, Universidad Nacional
                 Aut\'onoma de M\'exico, AP 70-264, Distrito Federal 04510,
	         M\'exico \\
             \(^2\) Departamento de F\'{\i}sica Te\'orica and IFIC, 
                  Centro Mixto Universidad de Valencia - CSIC. 
                  Universidad de Valencia, Burjassot-46100, Valencia, Spain
            }

\date{\today}


\begin{abstract}
 We give precise details to support that observations of gravitational
lensing at scales of  individual, groups and clusters of galaxies can
be understood in terms of  non-Newtonian gravitational interactions
with a relativistic structure compatible with the Einstein Equivalence
Principle.  This result is derived on very general grounds without knowing
the underlying structure of the gravitational field equations.  As such,
any developed gravitational theory built to deal with these astrophysical
scales needs to reproduce the obtained results of this article.
\end{abstract}

\maketitle

\section{Introduction}
\label{introduction}

  The first solid step towards a full development of a non-relativistic
theory of gravity was made by Newton in his \emph{Philosophi\ae Naturalis
Principia Mathematica} book~\citep{principia}.  The starting point of this
non-relativistic theory of gravity began with the third law of planetary
motion published by Kepler in his \emph{Harmonices Mundi}
book~\citep{kepler}.  For the
known 7 planets back then, this law represents a relation between the
mass of the sun \( M \), a planet's particular distance to the sun \(
r \) and the velocity \( v \) of a planet about the sun:
\( v \propto \left( M / r \right)^{1/2} \), for circular orbits. The
requirement of centripetal balance during the motion of planets yields:

\begin{equation}
  a =  - v^2 / r =  - G M / r^2,
\label{e000}
\end{equation}

\noindent where the proportionality factor \( G \) is Newton's
gravitational constant and the minus sign appears because of the
attractive nature of gravity.  The acceleration \( a \) produced by the
sun on a test planet is thus given by a force inversely proportional
to its separation from it and linearly depends on the sun's mass.
The right hand side of equation~\eqref{e000} is the simplest form of
the mathematical force of gravity introduced by Newton.

  In recent years, through dynamical observations of galaxies
\citep[e.g][and references therein]{famaey12}, dwarf
spheroidal galaxies~\citep[cf.][]{hernandez10}, globular
clusters~\citep{hernandez12b,hernandez13} and even wide open
binaries~\citep{hernandez12a}, it has became clear that Kepler's third
law appears not to hold in its classical form on these systems, but
rather requires a modification known as the Tully-Fisher law:

\begin{equation}
  v \propto M^{1/4},
\label{e001}
\end{equation}

\noindent where \( v \) represents the velocity (or dispersion velocity for
a dynamically pressure supported astrophysical system)
and \( M \) is the mass  (could be internal mass within a radius
r) of the system.  Similarly to Newton's approach, the requirement of 
centripetal balance means that the acceleration \( a \propto
v^2 / r \) at a distance \( r \) from the configuration's centre and so

\begin{equation}
   a = - G_\text{M} \frac{ M^{1/2}  }{ r },
\label{e002}
\end{equation}

\noindent where the  constant of proportionality has been written as \(
G_\text{M} \) and the minus sign has been introduced in order to manifest
the attractive nature of the gravitational force.    Equation~\eqref{e002}
can be seen as a motivation to suspect that a new theory of gravity needs
to be developed in these astrophysical systems, since its right hand
side represents a relation between the acceleration felt by a test body
of mass \( M \) at a distance \( r \).  In this sense, the proportionality
constant \( G_\text{M} \) can be seen as a new gravitational constant,
with dimensions of squared length over squared time by the square
root of mass.

  This means that, in the same way as \( G \) is
regarded as a fundamental constant of nature, \( G_\text{M} \) should
aspire to the same privileged status.  However, in order to gain
merits in that direction, \( G_\text{M} \) should play an essential
role in the description of relativistic phenomena on its corresponding
scales. Nonetheless, it should be noted that the construction of
equations~\eqref{e000} and~\eqref{e002} are completely independent,
since they both depend on different and unrelated data sets.  As such,
the constants \( G \) and \( G_\text{M} \) can safely be postulated
as independent.  Given this independence, one is allowed to think of
both as equally fundamental.

  Requiring gravity to be described by equation~\eqref{e000} at
some particular scales and behaving at some others according to
relation~\eqref{e002}, means that the scale invariance of gravity is
necessarily broken. One can postulate that at some astrophysical scales
gravity is Newtonian and requires modification at some others.  The scale
is not just a ``fixed'' distance scale. From the experimental astronomical
evidence mentioned above it follows that the modified regime of gravity
appears when the ratio of the mass of a given astrophysical system divided
by its characteristic radius is sufficiently small as compared to the
corresponding solar system value, which suggests that
the transition scale is dynamical rather a simple fixed length. A given
test particle sufficiently far away from a mass distribution is thus in
this modified regime of gravity.

  The  approach introduced above for the description of gravitational
phenomena departing from standard Newtonian gravity can be connected with
the simplest version of the Modified Newtonian Dynamics (MOND) formula
by replacing the constant \( G_\text{M} \) with a new constant \( a_0 \)
introduced by \citet{milgrom83a} with dimensions of acceleration through
the relation:

\begin{equation}
  a_0 := G_\text{M}^2 / G, 
\label{e002b}
\end{equation}

\noindent and so equation~\eqref{e002} can be written as:

\begin{equation}
   a = - \frac{ \left( a_0 G M \right)^{1/2}  }{ r }.
\label{e002a}
\end{equation}

\noindent Since Milgrom's acceleration constant \( a_0 \approx 1.2 \times
10^{-10} \, \mathrm{m} \, \mathrm{s}^{-2} \)~\citep{famaey12} it follows that:

\begin{equation}
  G_\text{M} \approx 8.94 \times 10^{-11} \, \mathrm{m}^2 \, \mathrm{s}^{-2} \,
    \mathrm{kg}^{-1/2}.
\label{e002a1}
\end{equation}

  In this regard it is quite important to notice that the formulation
of \citet{milgrom83a} describes a  modification on the dynamical
sector of Newton's second law and not on the particular form of the
gravitational force~\citep[cf.][]{milgrom06}.  This is quite evident from
the initial development of the theory, in which the requirement that
the squared of the acceleration \( a^2 \) proportional to the
Newtonian acceleration \( G M / r^2 \) implies flattening of rotation
curves in spiral galaxies.  In this relation, the proportionality constant
\( a_0 \) with dimensions of acceleration, is required to be a fundamental
quantity of nature.  By doing so,  the Tully-Fisher law is obtained
as a consequence of the proposed modification of dynamics.

  With the approach made here, it follows that the Tully-Fisher
law forces the construction of a full gravitational theory in systems
where Newtonian gravity does not work.  In its
simplest form, the developed theory must converge to
equation~\eqref{e002}.  As such, no need for modification of Newton's
second law needs to be introduced, since only a non-scale invariant
character for the gravitational law is directly inferred from the
observational data.

  In our view, the introduction of \( G_\text{M} \) as a fundamental
constant of gravity, rather than \( a_0 \) as a new fundamental
acceleration scale, sheds light onto the strategy to follow to unveil
the structure of the underlying theory. In fact, \( G_\text{M} \)
points towards a modification on the gravitational sector, whereas
\( a_0 \) could point towards a break down or possible extensions of
special relativity (due to the existence of a universal acceleration
scale, similarly as with the speed of light), with potentially dramatic
implications even in non-gravitational systems.

  If Newtonian gravity breaks at a certain scale, one can legitimately
wonder whether  the relativistic structure of gravitational interactions
remains valid or may require a full reformulation.  To explore these
aspects one should study not only the dynamics of slow massive particles
(e.g. equation~\eqref{e002}) but also the motion of relativistic particles
(such as photons) in astrophysical scenarios probing the gravitational
field in this new regime.  Being conservative, one may assume that
Einstein's insights on the geometrical interpretation of gravity remain
valid in this regime.  As such, it is perfectly reasonable to assume
that the Einstein Equivalence Principle remains valid, which implies
that test particles satisfy the geodesic equation:

\begin{equation}
  \frac{ \mathrm{d}^2 x^\alpha }{ \mathrm{d} s^2 }  + \Gamma^\alpha_{\ \mu\nu} 
    \frac{ \mathrm{d} x^{\mu} }{ \mathrm{d} s } \frac{ \mathrm{d} x^{\nu}
    }{ \mathrm{d} s } = 0,
\label{e003}
\end{equation}

\noindent where  \( \Gamma^{\alpha}_{\ \beta\eta} \) are the Christoffel
symbols and summation convention is used over repeated indices (Greek
indices vary from 0 to 3 and Latin ones from 1 to 3). The coordinates \(
x^\alpha = \left( ct, x, y,z \right) \) and the interval \( \mathrm{d} s^2
= g_{\mu\nu} \mathrm{d} x^\mu \mathrm{d} x^\nu \) for a metric tensor \(
g_{\mu\nu} \) and a velocity of light \( c\).

  By knowing Tully-Fisher's modification of Kepler's third
law~\eqref{e001} and the geodesic equation~\eqref{e003}, then at second
order perturbation, the bending of light is completely determined up to
a constant~\citep{will93}.  This is due to the fact that to this order of
approximation the motion of photons only depends on the non-relativistic
gravitational potential and a parameter \( \gamma = \text{const.} \),
which  measures the proportionality between the leading (second) order
corrections of the time and spatial metric components in isotropic coordinates.

  We show in this article that experimental data from astronomical
observations point us to show that only modifications of Kepler's third
law are necessary in order to reproduce observations of gravitational
lensing.  In this article we postulate the Einstein Equivalence Principle
to be valid and combine it with the non-relativistic gravitational
potential associated to the modified Kepler's third law (Tully-Fisher
law).  This approach only assumes that gravitation is a geometric
phenomenon and as such, we do not need to know the underlying set of
relativistic field equations to find that the corresponding predictions
for the bending of light are compatible with the observations of galaxies
and groups of galaxies.  This approach is parallel to the strategy
followed to understand the relativistic behaviour of gravity in the
solar system, where traditional Kepler's third law holds.

  Additionally, we interpret the results by \citet{hernandez12a} on
the failure of Kepler's third law for wide binary systems,  as a way
to test a key aspect of the mathematical structure of the underlying
theory of gravity, namely whether or not external boundary conditions
influence the internal dynamics of local gravitational systems, which
is sometimes referred to as an external field effect~\citep{famaey12}.
This effect means that for example, a gravitating system in the modified
Keplerian regime embedded on an external standard Newtonian (or Keplerian)
field, would behave in a Newtonian way.  \citet{hernandez12a} studied
orbits of wide binary stars  $\sim
1 M_{\odot}$ separated by \( \gtrsim 7000 \)~AU.  These bound objects
are embedded in our galaxy and are subject to its Newtonian gravity.
As such, if an external field effect occurs, then these objects would orbit
each other in a standard way, following Kepler's third law.
However, their analysis shows that a violation
of Kepler's third law occurs in these systems.  The large statistics and
precise astrometry to be obtained with the GAIA probe of the European
Space Agency in the near future, should provide a strong test for
the validity of Kepler's third law at scales yet to be explored.
Furthermore, we show that lensing observations strongly support
the validity of equation~\eqref{e003}, implying that the effects of
external gravitational fields can be removed by a suitable choice of
local coordinates (a freely falling frame).  To the light of these results, 
the idea of an external field effect appears as an artificial construction
(possibly related to the specific mathematical realisations of the theory).

  The article is organised as follows.  In
section~\ref{relativistic-extension} we discuss the simplest properties
that an extended theory of gravity must obey on systems where the
Tully-Fisher law is valid. In section~\ref{tully-fisher-relativistic}
we use gravitational lensing observations in individual, groups and
clusters of galaxies combined with the Tully-Fisher law to obtain
empirical relations for the metric coefficients of a spherically
symmetric space-time at second perturbation order.  Finally in
section~\ref{discussion} we summarise our main results and discuss them
in lights of future theoretical developments in the search for a complete
extended theory of gravity not requiring the use of dark matter in the
description of astrophysical phenomena

\section{Basic properties for a relativistic extensions of the Tully-Fisher
law}
\label{relativistic-extension}

  In order to seek for a complete set of fundamental observables of the
underlying  relativistic theory of gravity in the regime where
Kepler's third law requires modification in the form of a Tully-Fisher
law, let us analyse the behaviour of gravity in the weak
field limit, having in mind the behaviour of massive particles and photons.

  Consider a fixed point mass \( M \) at the centre of coordinates
generating a gravitational field.  The underlying 
space-time is thus static and its spherically symmetric metric line element 
\( \mathrm{d} s \) can be written in spherical Schwarzschild coordinates as:

\begin{equation}
  \mathrm{d}s^2 = g_{\mu\nu} \mathrm{d}x^\mu \mathrm{d} x^\nu  
    = g_{00} \, c^2 \mathrm{d}t^2 + g_{11}\mathrm{d}r^2
	-r^2 \mathrm{d} \Omega^2.
\label{metric}
\end{equation}

\noindent  In the previous equation and in what follows the space-time
coordinates \( (x^0,x^1,x^2,x^3) = (ct,r,\theta,\varphi) \), where
\(t \) represents time, \( r \) the radial coordinate and the polar and
azimuthal angles are given by \( \theta \) and \( \varphi \) respectively.
The angular displacement \( \mathrm{d} \Omega^2 := \mathrm{d}\theta^2
+ \sin^2\theta \, \mathrm{d}\varphi^2 \).  The symmetry of the problem
means that the unknown metric components \( g_{00} \) and \( g_{11} \) are
functions that depend on the radial coordinate \( r \) only.  In the limit
where the speed of light \( c \rightarrow \infty \), the radial component
of the geodesic equation~\eqref{e003} in these coordinates is given by:

\begin{equation}
  \frac{1}{c^2} \frac{\text{d}^2 r}{\text{d} t^2} =
   \frac{1}{2} g^{11} \frac{ \partial g_{00}  }{ \partial r }.
\label{radialgeodesic}
\end{equation}

\noindent The previous
relation  holds since we have used the following approximation for the weak
field limit: \( \text{d}s = c \ \text{d}t \), and so, due to the fact that
the velocity \(v \ll c\) then \( v^i  \ll \text{d}x^0 / \text{d}t\) with 
\( v^i := \left( \mathrm{d}r / \mathrm{d}t,\ r \mathrm{d} \theta / \mathrm{d}t,\
r\sin \theta \, \mathrm{d} \varphi / \mathrm{d}t \right) \).

  The lowest perturbation order of equation~\eqref{radialgeodesic} is
obtained when its left-hand side is of order \(v^2 / c^2 \) and when its
right-hand side is of order \(  g_{00} = 1 + \phi / c^2 \)~\citep{will93}.
Both are orders \( \mathcal{O}(1/c^2) \) of the underlying theory,
or simply \( \mathcal{O}(2) \).

  In this weak-field slow-motion approximation, a particle bounded to a circular
orbit about the mass \( M \) experiences a centrifugal radial acceleration
given by:

\begin{equation}
  \frac{\text{d}^2 r}{\text{d} t^2} =  \frac{v^2}{r},
\label{centrifugal}
\end{equation}

\noindent for a circular tangential velocity \( v \).

 The motion of material and light particles at this lowest perturbation
order is such that the metric components are given by~\citep{will93,will06}:

\begin{equation}
\begin{split}
	g_{00} =& {}^{(0)}g_{00} + {}^{(2)}g_{00} + \mathcal{O}(4)= 1 + {}^{(2)}g_{00}
	  + \mathcal{O}(4),  \\
	g_{11} =& {}^{(0)}g_{11} + {}^{(2)}g_{11} + \mathcal{O}(4)
	  = - 1 + {}^{(2)}g_{11} + \mathcal{O}(4),  \\
	g_{22} =&  -r^2,  \\
	g_{33} =& - r^2 \sin^2 \theta,
\end{split}
\label{metric-perturbed}
\end{equation}

\noindent where the superscript \( (p) \) denotes the order \(
\mathcal{O}(p) \) at which a particular quantity is approximated. 
The non-relativistic potential \( \phi \) is defined as 
\citep[e.g][]{daufields,will93,MTW}:

\begin{equation}
  {}^{(2)}g_{00} = \frac{ 2 \phi  }{ c^2 }. 
\label{potential}
\end{equation}

From equations~\eqref{metric-perturbed} it follows that the contravariant
metric components are given by:

\begin{equation}
\begin{split}
	g^{00} =& {}^{(0)}g^{00} + {}^{(2)}g^{00} + \mathcal{O}(4)
	  = 1 - {}^{(2)}g_{00} + \mathcal{O}(4),  \\
	g^{11} =& {}^{(0)}g^{11} + {}^{(2)}g^{11} + \mathcal{O}(4)
	  = - 1 - {}^{(2)}g_{11} + \mathcal{O}(4),  \\
	g^{22} =& - 1/r^2, 	 \\
	g^{33} =& - 1 / r^2 \sin^2 \theta.
\end{split}
\label{metric:exp-up}
\end{equation}

  At this level of approximation, the motion of non-relativistic massive
particles only requires knowledge of the metric component \( {^{(2)}}g_{00}
\).  The motion of photons is fully determined by additionally knowing 
\( ^{{(2)}}g_{11} \)~\citep[cf.][]{will93}.

\section{Tully-Fisher's relativistic corrections}
\label{tully-fisher-relativistic}

  Let us take the radial component~\eqref{radialgeodesic} of the geodesic
equations~\eqref{e003} at the lowest relativistic perturbation order.
In this limit, the  rotation curve for test particles bound to a circular
orbit about a mass \(M\) with circular velocity \( v \) is given by
equation~\eqref{centrifugal} and so:

\begin{equation}
   \frac{v^{2} }{c^{2} r} = \frac{1}{2} 
      { \frac{ \partial \, {^{(2)}}g_{00} }{ \partial r } }.
\label{v:metric}
\end{equation}

\noindent Since we are interested in the behaviour of particles where
the modified Kepler's third law (or Tully-Fisher law) holds,
equation~\eqref{e001} can be written as:

\begin{equation}
  v = G_\text{M}^{1/2} \ M^{1/4}.
\label{tully-fisher-law}
\end{equation}

 Substitution of this equation on relation~\eqref{v:metric} yields:

\begin{equation}
  \frac{ \partial \, ^{(2)}g_{00} }{ \partial r }  = 
    - \frac{2}{r} \left( \frac{v}{c} \right)^2 
     = - \frac{2 G_\text{M} M^{1/2} }{ c^{2} r } ,
\end{equation}

\noindent which has a direct analytical solution:

\begin{equation}
  {^{(2)}}g_{00} = - 2 \left( \frac{v}{c} \right)^2
    \ln \left( \frac{r}{r_\star} \right)  
  = - \frac{ 2  G_\text{M} M^{1/2} }{ c^{2} } \ln
    \left( \frac{r}{r_\star} \right),
\label{g00:empirical}
\end{equation}

\noindent where \( r_\star \) is an arbitrary length.

 Having obtained the  \( {^{(2)}}g_{00} \) component, which determines the
non-relativistic motion of massive particles, we now proceed to obtain
the \( {^{(2)}}g_{11} \) component. In the literature it is customary
to define a new scalar potential \( \psi \) as:

\begin{equation}
 {^{(2)}}g_{11} = - \frac{ 2 \psi}{c^2},
\label{anotherpotential}
\end{equation}

\noindent in complete analogy with equation~\eqref{potential}.  The
introduction of this potential can be justified considering a more general
scenario.  Without requiring spherical symmetry, the
spatial part of the metric can be written as \( g_{ik} \mathrm{d}x^i
\mathrm{d}x^k \), with \( {^{(0)}}g_{kl} = \delta_{kl} \) being the
Minkowskian part. The second order perturbation corrections of \( g_{kl} \)
could in principle involve other potentials (and not only \( \phi \) or \(
\psi \)).  By a suitable choice of coordinates, one can get rid of the
anisotropic contributions at the same perturbation order, which turns \(
g_{kl} \) into a diagonal form.  Given the isotropy of space, there is no
preferred direction and so \( {^{(2)}}g_{ik} \propto \delta_{ik} \).
It is natural to expect that the leading order \(
\mathcal{O}(2) \) correction must be of the same order of magnitude as the
gravitational potential \( \phi \).  Accordingly \( g_{kl} = \left( 1 +
2 \gamma \phi / c^2  \right) \delta_{kl} \), where \( \gamma \) is a
proportionality constant,  and so

\begin{equation}
  \mathrm{d} s^2 = g_{00} \mathrm{d} t  - \left( 1 +
     2 \gamma \phi / c^2  \right) \delta_{kl}
  \mathrm{d}x^k \mathrm{d}x^l.
\label{isotropic}
\end{equation}

  Since spherical Schwarzschild coordinates are widely used in
astrophysical literature, let us calculate the metric component 
\( g_{11} \) in such coordinates.  The conversion is straightforward since

\begin{equation}
  g_{11} \mathrm{d} r^2 + r^2 \mathrm{d} \Omega^2 = \left( 1 +
  2 \gamma \phi / c^2  \right) \left(
  \mathrm{d}\tilde{r}^2  + \tilde{r}^2 \mathrm{d} \Omega^2 \right),
\label{conversion}
\end{equation}

\noindent for spherical isotropic coordinates 
(\( c t,\tilde{r},\theta,\varphi\)).

\noindent Using equations~\eqref{g00:empirical} and~\eqref{potential} it
follows that 

\begin{gather}
   r = \tilde{r} \left[ 1 - \gamma \left( \frac{ G_\text{M} M^{1/2} }{ c^2 } 
     \right) \ln\left( \frac{ r }{ r_\star } \right) \right], 
     					\label{t01} \\
   \intertext{and so,}
   \mathrm{d} r = \mathrm{d} \tilde{r} \left[ 1 - \frac{ G_\text{M} M^{1/2} 
     }{ c^2 } \ln\left( \frac{ \tilde{r} }{ r_\star} \right) - 
     \frac{ G_\text{M} M^{1/2} }{ c^2 } \right],
\label{t02}
\end{gather}

\noindent at perturbation order \( \mathcal{O}(2) \).  This means that:

\begin{gather}
   \psi = - \gamma G_\text{M} M^{1/2},
   					\label{eq06} \\
  \intertext{which yields:}
    {^{(2)}}g_{11} = - \frac{ 2 \gamma G_\text{M} M^{1/2} }{ c^2 }.
\label{eq07}
\end{gather}

  As pointed out by \citet{mendoza13}, recent
observations have shown that gravitational lensing on
individual~\citep{gavazzi07,koopmans06,barnabe11,suyu12,dutton11},
groups~\citep{more11} and clusters of galaxies~\citep{newman09,limousin07}
can be modelled with the standard Schwarzschild solution of general
relativity, assuming the existence of a total dark plus baryonic
isothermal halo, where the Tully-Fisher law holds for the baryonic
matter. As such, the bending angle of light can be calculated using
the standard lensing equation, finding that it does not depend on the
impact parameter and scales with the square root of the total baryonic
mass.  Under a modified theory of gravity scheme with no dark matter component, 
the bending angle has the same value and so, since the time metric
component \( {}^{(2)}g_{00} \) is known from the Tully-Fisher law as shown
in equation~\eqref{g00:empirical}, then a reconstruction of the spatial  metric component \( {}^{(2)}g_{11} \)can be made.  The final result is that
the gravitational potentials on this modified regime of gravity have the
following values:

\begin{equation}
 \phi = -  G_\text{M} M^{1/2} \ln \left( \frac{r}{r_\star} \right),
  \quad \text{and}  \quad
  \psi = - G_\text{M} M^{1/2}.
\label{resultsmendoza} 
\end{equation}

  Comparison of this last equation with relations~\eqref{eq06}
and~\eqref{g00:empirical} yields:

\begin{equation}
  \gamma = 1.
\label{gamma}
\end{equation}

\noindent We thus conclude that the relativistic structure of the
underlying theory of gravity at individual, groups and cluster of galaxies
scales is compatible with that found in the solar system, where \( \gamma
= 1 \).  The difference however, lies on the fact that the gravitational
potential \( \phi \) appearing in equation~\eqref{isotropic} is not the one
associated with Kepler's third law, but the one inferred from the
Tully-Fisher law.

\section{Discussion}
\label{discussion}

A number of independent astrophysical observations strongly
support the view that the scale invariance of Newtonian gravity breaks down
at scales that depend on the mass and characteristic sizes of the systems
involved.  

  In the non-relativistic weak field limit of approximation, the behaviour of
gravity is Newtonian, and the full relativistic theory that describes
objects in this regime is general relativity.  At the very weak field limit
of approximation, sufficiently far from the masses that produce the
gravitational field, Kepler's third law is modified through the
Tully-Fisher law and the underlying relativistic theory in this regime is
so far unknown.

  We have explored some relativistic properties of this modified regime of
gravity at the weak field limit of approximation, assuming that gravity
is a geometrical phenomenon and that  the Einstein Equivalence Principle
holds.  This is sufficient to build a model independent approach of
the relativistic regime at second perturbation order \( \mathcal{O}(2)
\), in complete analogy to the one used at solar system scales where
the dynamics are compatible with Einstein's general relativity.
Using this modified Kepler's third law and lensing observations for
individual, groups and clusters of galaxies, and interpreted in the way
\citet{mendoza13} did, we have shown that this relativistic approach
is in exact compatibility with these observations.  In isotropic
coordinates, the non-relativistic gravitational potentials \( \phi \)
and \( \psi \) defined in equations~\eqref{potential} and~\eqref{eq06}
are proportional to each other, i.e. \( \phi \propto \gamma \psi \),
where \( \gamma \) is a Parametrised Post Newtonian Parameter in this
new regime of gravity where Kepler's third law does not hold.  As shown
in this article, lensing observations require \( \gamma = 1 \), as in
general relativity.  Note that in spherical Schwarzschild coordinates both
gravitational scalar potentials differ from each other as its evident
from equations~\eqref{resultsmendoza}.  For the case of Einstein's general
relativity these potentials are coincidentally equal not only in isotropic
coordinates, but also in spherical Schwarzschild coordinates.

  As shown in this article, the bending of light in regions where the
Tully-Fisher law is satisfied can be predicted  at second order
perturbation without knowledge of the underlying relativistic theory of
gravity. This is fully consistent with Einstein's view on the geometrical
nature of space-time and relativistic motion.  Any viable extended relativistic
theory of gravity should be in agreement with the light bending predictions
discussed here.  The proposal by \citet{bernal11} with lensing applications
detailed in \citet{mendoza13} is an example of such kind of theory.

\section{Acknowledgements}
\label{acknowledgements}
This work was supported by a DGAPA-UNAM grant (IN111513-3), by the
Programa de Cooperacion Cientifica UNAM-CSIC, by the Spanish grants
FIS2011-29813-C02-02 and i-LINK0780 (CSIC),  and by the JAE-doc program
of the Spanish Research Council (CSIC). SM acknowledges economic support
from CONACyT (26344) and the hospitality of the Instituto de F\'{\i}sica
Corpuscular (Universidad de Valencia \& CSIC).  GJO acknowledges the
hospitality of the Instituto de Astronom\'{\i}a at the Universidad
Nacional Aut\'onoma de M\'exico where the first ideas of this article
were first developed.


\bibliographystyle{aipnum4-1}
\bibliography{PPnN}

\begin{thebibliography}{23}%
\makeatletter
\providecommand \@ifxundefined [1]{%
 \@ifx{#1\undefined}
}%
\providecommand \@ifnum [1]{%
 \ifnum #1\expandafter \@firstoftwo
 \else \expandafter \@secondoftwo
 \fi
}%
\providecommand \@ifx [1]{%
 \ifx #1\expandafter \@firstoftwo
 \else \expandafter \@secondoftwo
 \fi
}%
\providecommand \natexlab [1]{#1}%
\providecommand \enquote  [1]{``#1''}%
\providecommand \bibnamefont  [1]{#1}%
\providecommand \bibfnamefont [1]{#1}%
\providecommand \citenamefont [1]{#1}%
\providecommand \href@noop [0]{\@secondoftwo}%
\providecommand \href [0]{\begingroup \@sanitize@url \@href}%
\providecommand \@href[1]{\@@startlink{#1}\@@href}%
\providecommand \@@href[1]{\endgroup#1\@@endlink}%
\providecommand \@sanitize@url [0]{\catcode `\\12\catcode `\$12\catcode
  `\&12\catcode `\#12\catcode `\^12\catcode `\_12\catcode `\%12\relax}%
\providecommand \@@startlink[1]{}%
\providecommand \@@endlink[0]{}%
\providecommand \url  [0]{\begingroup\@sanitize@url \@url }%
\providecommand \@url [1]{\endgroup\@href {#1}{\urlprefix }}%
\providecommand \urlprefix  [0]{URL }%
\providecommand \Eprint [0]{\href }%
\providecommand \doibase [0]{http://dx.doi.org/}%
\providecommand \selectlanguage [0]{\@gobble}%
\providecommand \bibinfo  [0]{\@secondoftwo}%
\providecommand \bibfield  [0]{\@secondoftwo}%
\providecommand \translation [1]{[#1]}%
\providecommand \BibitemOpen [0]{}%
\providecommand \bibitemStop [0]{}%
\providecommand \bibitemNoStop [0]{.\EOS\space}%
\providecommand \EOS [0]{\spacefactor3000\relax}%
\providecommand \BibitemShut  [1]{\csname bibitem#1\endcsname}%
\let\auto@bib@innerbib\@empty
\bibitem [{\citenamefont {Newton}(1729)}]{principia}%
  \BibitemOpen
  \bibfield  {author} {\bibinfo {author} {\bibfnamefont {I.}~\bibnamefont
  {Newton}},\ }\href {http://books.google.es/books?id=Tm0FAAAAQAAJ} {\emph
  {\bibinfo {title} {The Mathematical Principles of Natural Philosophy}}},\
  \bibinfo {edition} {3rd}\ ed.,\ The Mathematical Principles of Natural
  Philosophy\ (\bibinfo  {publisher} {B. Motte},\ \bibinfo {year}
  {1729})\BibitemShut {NoStop}%
\bibitem [{\citenamefont {{Kepler}}, \citenamefont {{Ptolemaeus}},\ and\
  \citenamefont {{Fludd}}(1619)}]{kepler}%
  \BibitemOpen
  \bibfield  {author} {\bibinfo {author} {\bibfnamefont {J.}~\bibnamefont
  {{Kepler}}}, \bibinfo {author} {\bibfnamefont {C.}~\bibnamefont
  {{Ptolemaeus}}}, \ and\ \bibinfo {author} {\bibfnamefont {R.}~\bibnamefont
  {{Fludd}}},\ }\href@noop {} {\emph {\bibinfo {title} {Lincii Austriae,
  sumptibus G.~Tampachii, excudebat I.~Plancvs, 1619.}}}\ (\bibinfo {year}
  {1619})\BibitemShut {NoStop}%
\bibitem [{\citenamefont {{Famaey}}\ and\ \citenamefont
  {{McGaugh}}(2012)}]{famaey12}%
  \BibitemOpen
  \bibfield  {author} {\bibinfo {author} {\bibfnamefont {B.}~\bibnamefont
  {{Famaey}}}\ and\ \bibinfo {author} {\bibfnamefont {S.~S.}\ \bibnamefont
  {{McGaugh}}},\ }\href {\doibase 10.12942/lrr-2012-10} {\bibfield  {journal}
  {\bibinfo  {journal} {Living Reviews in Relativity}\ }\textbf {\bibinfo
  {volume} {15}},\ \bibinfo {pages} {10} (\bibinfo {year} {2012})},\ \Eprint
  {http://arxiv.org/abs/1112.3960} {arXiv:1112.3960 [astro-ph.CO]} \BibitemShut
  {NoStop}%
\bibitem [{\citenamefont {{Hernandez}}\ \emph {et~al.}(2010)\citenamefont
  {{Hernandez}}, \citenamefont {{Mendoza}}, \citenamefont {{Suarez}},\ and\
  \citenamefont {{Bernal}}}]{hernandez10}%
  \BibitemOpen
  \bibfield  {author} {\bibinfo {author} {\bibfnamefont {X.}~\bibnamefont
  {{Hernandez}}}, \bibinfo {author} {\bibfnamefont {S.}~\bibnamefont
  {{Mendoza}}}, \bibinfo {author} {\bibfnamefont {T.}~\bibnamefont {{Suarez}}},
  \ and\ \bibinfo {author} {\bibfnamefont {T.}~\bibnamefont {{Bernal}}},\
  }\href {\doibase 10.1051/0004-6361/200913301} {\bibfield  {journal} {\bibinfo
   {journal} {\aap}\ }\textbf {\bibinfo {volume} {514}},\ \bibinfo {eid} {A101}
  (\bibinfo {year} {2010})},\ \Eprint {http://arxiv.org/abs/0904.1434}
  {arXiv:0904.1434 [astro-ph.GA]} \BibitemShut {NoStop}%
\bibitem [{\citenamefont {{Hernandez}}\ and\ \citenamefont
  {{Jim{\'e}nez}}(2012)}]{hernandez12b}%
  \BibitemOpen
  \bibfield  {author} {\bibinfo {author} {\bibfnamefont {X.}~\bibnamefont
  {{Hernandez}}}\ and\ \bibinfo {author} {\bibfnamefont {M.~A.}\ \bibnamefont
  {{Jim{\'e}nez}}},\ }\href {\doibase 10.1088/0004-637X/750/1/9} {\bibfield
  {journal} {\bibinfo  {journal} {\apj}\ }\textbf {\bibinfo {volume} {750}},\
  \bibinfo {eid} {9} (\bibinfo {year} {2012})},\ \Eprint
  {http://arxiv.org/abs/1108.4021} {arXiv:1108.4021 [astro-ph.CO]} \BibitemShut
  {NoStop}%
\bibitem [{\citenamefont {{Hernandez}}, \citenamefont {{Jim{\'e}nez}},\ and\
  \citenamefont {{Allen}}(2013)}]{hernandez13}%
  \BibitemOpen
  \bibfield  {author} {\bibinfo {author} {\bibfnamefont {X.}~\bibnamefont
  {{Hernandez}}}, \bibinfo {author} {\bibfnamefont {M.~A.}\ \bibnamefont
  {{Jim{\'e}nez}}}, \ and\ \bibinfo {author} {\bibfnamefont {C.}~\bibnamefont
  {{Allen}}},\ }\href {\doibase 10.1093/mnras/sts263} {\bibfield  {journal}
  {\bibinfo  {journal} {\mnras}\ }\textbf {\bibinfo {volume} {428}},\ \bibinfo
  {pages} {3196} (\bibinfo {year} {2013})},\ \Eprint
  {http://arxiv.org/abs/1206.5024} {arXiv:1206.5024 [astro-ph.GA]} \BibitemShut
  {NoStop}%
\bibitem [{\citenamefont {{Hernandez}}, \citenamefont {{Jim{\'e}nez}},\ and\
  \citenamefont {{Allen}}(2012)}]{hernandez12a}%
  \BibitemOpen
  \bibfield  {author} {\bibinfo {author} {\bibfnamefont {X.}~\bibnamefont
  {{Hernandez}}}, \bibinfo {author} {\bibfnamefont {M.~A.}\ \bibnamefont
  {{Jim{\'e}nez}}}, \ and\ \bibinfo {author} {\bibfnamefont {C.}~\bibnamefont
  {{Allen}}},\ }\href {\doibase 10.1140/epjc/s10052-012-1884-6} {\bibfield
  {journal} {\bibinfo  {journal} {European Physical Journal C}\ }\textbf
  {\bibinfo {volume} {72}},\ \bibinfo {pages} {1884} (\bibinfo {year}
  {2012})},\ \Eprint {http://arxiv.org/abs/1105.1873} {arXiv:1105.1873
  [astro-ph.GA]} \BibitemShut {NoStop}%
\bibitem [{\citenamefont {{Milgrom}}(1983)}]{milgrom83a}%
  \BibitemOpen
  \bibfield  {author} {\bibinfo {author} {\bibfnamefont {M.}~\bibnamefont
  {{Milgrom}}},\ }\href {\doibase 10.1086/161130} {\bibfield  {journal}
  {\bibinfo  {journal} {\apj}\ }\textbf {\bibinfo {volume} {270}},\ \bibinfo
  {pages} {365} (\bibinfo {year} {1983})}\BibitemShut {NoStop}%
\bibitem [{\citenamefont {{Milgrom}}(2006)}]{milgrom06}%
  \BibitemOpen
  \bibfield  {author} {\bibinfo {author} {\bibfnamefont {M.}~\bibnamefont
  {{Milgrom}}},\ }in\ \href {\doibase 10.1051/eas:2006074} {\emph {\bibinfo
  {booktitle} {EAS Publications Series}}},\ \bibinfo {series} {EAS Publications
  Series}, Vol.~\bibinfo {volume} {20},\ \bibinfo {editor} {edited by\ \bibinfo
  {editor} {\bibfnamefont {G.~A.}\ \bibnamefont {{Mamon}}}, \bibinfo {editor}
  {\bibfnamefont {F.}~\bibnamefont {{Combes}}}, \bibinfo {editor}
  {\bibfnamefont {C.}~\bibnamefont {{Deffayet}}}, \ and\ \bibinfo {editor}
  {\bibfnamefont {B.}~\bibnamefont {{Fort}}}}\ (\bibinfo {year} {2006})\ pp.\
  \bibinfo {pages} {217--224},\ \Eprint
  {http://arxiv.org/abs/arXiv:astro-ph/0510117} {arXiv:astro-ph/0510117}
  \BibitemShut {NoStop}%
\bibitem [{\citenamefont {{Will}}(1993)}]{will93}%
  \BibitemOpen
  \bibfield  {author} {\bibinfo {author} {\bibfnamefont {C.~M.}\ \bibnamefont
  {{Will}}},\ }\href@noop {} {\emph {\bibinfo {title} {Theory and Experiment in
  Gravitational Physics, by Clifford M.~Will, pp.~396.~ISBN
  0521439736.~Cambridge, UK: Cambridge University Press, March 1993.}}}\
  (\bibinfo  {publisher} {Cambridge University Press},\ \bibinfo {year}
  {1993})\BibitemShut {NoStop}%
\bibitem [{\citenamefont {{Will}}(2006)}]{will06}%
  \BibitemOpen
  \bibfield  {author} {\bibinfo {author} {\bibfnamefont {C.~M.}\ \bibnamefont
  {{Will}}},\ }\href@noop {} {\bibfield  {journal} {\bibinfo  {journal} {Living
  Reviews in Relativity}\ }\textbf {\bibinfo {volume} {9}},\ \bibinfo {pages}
  {3} (\bibinfo {year} {2006})},\ \Eprint {http://arxiv.org/abs/0510072}
  {arXiv:0510072 [gr-qc]} \BibitemShut {NoStop}%
\bibitem [{\citenamefont {Landau}\ and\ \citenamefont
  {Lifshitz}(1975)}]{daufields}%
  \BibitemOpen
  \bibfield  {author} {\bibinfo {author} {\bibfnamefont {L.}~\bibnamefont
  {Landau}}\ and\ \bibinfo {author} {\bibfnamefont {E.}~\bibnamefont
  {Lifshitz}},\ }\href@noop {} {\emph {\bibinfo {title} {The classical theory
  of fields}}},\ Course of theoretical physics\ (\bibinfo  {publisher}
  {Butterworth Heinemann},\ \bibinfo {year} {1975})\BibitemShut {NoStop}%
\bibitem [{\citenamefont {Misner}, \citenamefont {Thorne},\ and\ \citenamefont
  {Wheeler}(1973)}]{MTW}%
  \BibitemOpen
  \bibfield  {author} {\bibinfo {author} {\bibfnamefont {C.}~\bibnamefont
  {Misner}}, \bibinfo {author} {\bibfnamefont {K.}~\bibnamefont {Thorne}}, \
  and\ \bibinfo {author} {\bibfnamefont {J.}~\bibnamefont {Wheeler}},\ }\href
  {http://books.google.es/books?id=ExAbAQAAIAAJ} {\emph {\bibinfo {title}
  {Gravitation}}},\ Physics Series\ (\bibinfo  {publisher} {W. H. Freeman},\
  \bibinfo {year} {1973})\BibitemShut {NoStop}%
\bibitem [{\citenamefont {{Mendoza}}\ \emph {et~al.}(2013)\citenamefont
  {{Mendoza}}, \citenamefont {{Bernal}}, \citenamefont {{Hernandez}},
  \citenamefont {{Hidalgo}},\ and\ \citenamefont {{Torres}}}]{mendoza13}%
  \BibitemOpen
  \bibfield  {author} {\bibinfo {author} {\bibfnamefont {S.}~\bibnamefont
  {{Mendoza}}}, \bibinfo {author} {\bibfnamefont {T.}~\bibnamefont {{Bernal}}},
  \bibinfo {author} {\bibfnamefont {X.}~\bibnamefont {{Hernandez}}}, \bibinfo
  {author} {\bibfnamefont {J.~C.}\ \bibnamefont {{Hidalgo}}}, \ and\ \bibinfo
  {author} {\bibfnamefont {L.~A.}\ \bibnamefont {{Torres}}},\ }\href {\doibase
  10.1093/mnras/stt752} {\bibfield  {journal} {\bibinfo  {journal} {\mnras}\
  }\textbf {\bibinfo {volume} {433}},\ \bibinfo {pages} {1802} (\bibinfo {year}
  {2013})},\ \Eprint {http://arxiv.org/abs/1208.6241} {arXiv:1208.6241
  [astro-ph.CO]} \BibitemShut {NoStop}%
\bibitem [{\citenamefont {{Gavazzi}}\ \emph {et~al.}(2007)\citenamefont
  {{Gavazzi}}, \citenamefont {{Treu}}, \citenamefont {{Rhodes}}, \citenamefont
  {{Koopmans}}, \citenamefont {{Bolton}}, \citenamefont {{Burles}},
  \citenamefont {{Massey}},\ and\ \citenamefont {{Moustakas}}}]{gavazzi07}%
  \BibitemOpen
  \bibfield  {author} {\bibinfo {author} {\bibfnamefont {R.}~\bibnamefont
  {{Gavazzi}}}, \bibinfo {author} {\bibfnamefont {T.}~\bibnamefont {{Treu}}},
  \bibinfo {author} {\bibfnamefont {J.~D.}\ \bibnamefont {{Rhodes}}}, \bibinfo
  {author} {\bibfnamefont {L.~V.~E.}\ \bibnamefont {{Koopmans}}}, \bibinfo
  {author} {\bibfnamefont {A.~S.}\ \bibnamefont {{Bolton}}}, \bibinfo {author}
  {\bibfnamefont {S.}~\bibnamefont {{Burles}}}, \bibinfo {author}
  {\bibfnamefont {R.~J.}\ \bibnamefont {{Massey}}}, \ and\ \bibinfo {author}
  {\bibfnamefont {L.~A.}\ \bibnamefont {{Moustakas}}},\ }\href {\doibase
  10.1086/519237} {\bibfield  {journal} {\bibinfo  {journal} {\apj}\ }\textbf
  {\bibinfo {volume} {667}},\ \bibinfo {pages} {176} (\bibinfo {year}
  {2007})},\ \Eprint {http://arxiv.org/abs/arXiv:astro-ph/0701589}
  {arXiv:astro-ph/0701589} \BibitemShut {NoStop}%
\bibitem [{\citenamefont {{Koopmans}}\ \emph {et~al.}(2006)\citenamefont
  {{Koopmans}}, \citenamefont {{Treu}}, \citenamefont {{Bolton}}, \citenamefont
  {{Burles}},\ and\ \citenamefont {{Moustakas}}}]{koopmans06}%
  \BibitemOpen
  \bibfield  {author} {\bibinfo {author} {\bibfnamefont {L.~V.~E.}\
  \bibnamefont {{Koopmans}}}, \bibinfo {author} {\bibfnamefont
  {T.}~\bibnamefont {{Treu}}}, \bibinfo {author} {\bibfnamefont {A.~S.}\
  \bibnamefont {{Bolton}}}, \bibinfo {author} {\bibfnamefont {S.}~\bibnamefont
  {{Burles}}}, \ and\ \bibinfo {author} {\bibfnamefont {L.~A.}\ \bibnamefont
  {{Moustakas}}},\ }\href {\doibase 10.1086/505696} {\bibfield  {journal}
  {\bibinfo  {journal} {\apj}\ }\textbf {\bibinfo {volume} {649}},\ \bibinfo
  {pages} {599} (\bibinfo {year} {2006})},\ \Eprint
  {http://arxiv.org/abs/arXiv:astro-ph/0601628} {arXiv:astro-ph/0601628}
  \BibitemShut {NoStop}%
\bibitem [{\citenamefont {{Barnab{\`e}}}\ \emph {et~al.}(2011)\citenamefont
  {{Barnab{\`e}}}, \citenamefont {{Czoske}}, \citenamefont {{Koopmans}},
  \citenamefont {{Treu}},\ and\ \citenamefont {{Bolton}}}]{barnabe11}%
  \BibitemOpen
  \bibfield  {author} {\bibinfo {author} {\bibfnamefont {M.}~\bibnamefont
  {{Barnab{\`e}}}}, \bibinfo {author} {\bibfnamefont {O.}~\bibnamefont
  {{Czoske}}}, \bibinfo {author} {\bibfnamefont {L.~V.~E.}\ \bibnamefont
  {{Koopmans}}}, \bibinfo {author} {\bibfnamefont {T.}~\bibnamefont {{Treu}}},
  \ and\ \bibinfo {author} {\bibfnamefont {A.~S.}\ \bibnamefont {{Bolton}}},\
  }\href {\doibase 10.1111/j.1365-2966.2011.18842.x} {\bibfield  {journal}
  {\bibinfo  {journal} {\mnras}\ }\textbf {\bibinfo {volume} {415}},\ \bibinfo
  {pages} {2215} (\bibinfo {year} {2011})},\ \Eprint
  {http://arxiv.org/abs/1102.2261} {arXiv:1102.2261 [astro-ph.CO]} \BibitemShut
  {NoStop}%
\bibitem [{\citenamefont {{Suyu}}\ \emph {et~al.}(2012)\citenamefont {{Suyu}},
  \citenamefont {{Hensel}}, \citenamefont {{McKean}}, \citenamefont
  {{Fassnacht}}, \citenamefont {{Treu}}, \citenamefont {{Halkola}},
  \citenamefont {{Norbury}}, \citenamefont {{Jackson}}, \citenamefont
  {{Schneider}}, \citenamefont {{Thompson}}, \citenamefont {{Auger}},
  \citenamefont {{Koopmans}},\ and\ \citenamefont {{Matthews}}}]{suyu12}%
  \BibitemOpen
  \bibfield  {author} {\bibinfo {author} {\bibfnamefont {S.~H.}\ \bibnamefont
  {{Suyu}}}, \bibinfo {author} {\bibfnamefont {S.~W.}\ \bibnamefont
  {{Hensel}}}, \bibinfo {author} {\bibfnamefont {J.~P.}\ \bibnamefont
  {{McKean}}}, \bibinfo {author} {\bibfnamefont {C.~D.}\ \bibnamefont
  {{Fassnacht}}}, \bibinfo {author} {\bibfnamefont {T.}~\bibnamefont {{Treu}}},
  \bibinfo {author} {\bibfnamefont {A.}~\bibnamefont {{Halkola}}}, \bibinfo
  {author} {\bibfnamefont {M.}~\bibnamefont {{Norbury}}}, \bibinfo {author}
  {\bibfnamefont {N.}~\bibnamefont {{Jackson}}}, \bibinfo {author}
  {\bibfnamefont {P.}~\bibnamefont {{Schneider}}}, \bibinfo {author}
  {\bibfnamefont {D.}~\bibnamefont {{Thompson}}}, \bibinfo {author}
  {\bibfnamefont {M.~W.}\ \bibnamefont {{Auger}}}, \bibinfo {author}
  {\bibfnamefont {L.~V.~E.}\ \bibnamefont {{Koopmans}}}, \ and\ \bibinfo
  {author} {\bibfnamefont {K.}~\bibnamefont {{Matthews}}},\ }\href {\doibase
  10.1088/0004-637X/750/1/10} {\bibfield  {journal} {\bibinfo  {journal}
  {\apj}\ }\textbf {\bibinfo {volume} {750}},\ \bibinfo {eid} {10} (\bibinfo
  {year} {2012})},\ \Eprint {http://arxiv.org/abs/1110.2536} {arXiv:1110.2536
  [astro-ph.CO]} \BibitemShut {NoStop}%
\bibitem [{\citenamefont {{Dutton}}\ \emph {et~al.}(2011)\citenamefont
  {{Dutton}}, \citenamefont {{Brewer}}, \citenamefont {{Marshall}},
  \citenamefont {{Auger}}, \citenamefont {{Treu}}, \citenamefont {{Koo}},
  \citenamefont {{Bolton}}, \citenamefont {{Holden}},\ and\ \citenamefont
  {{Koopmans}}}]{dutton11}%
  \BibitemOpen
  \bibfield  {author} {\bibinfo {author} {\bibfnamefont {A.~A.}\ \bibnamefont
  {{Dutton}}}, \bibinfo {author} {\bibfnamefont {B.~J.}\ \bibnamefont
  {{Brewer}}}, \bibinfo {author} {\bibfnamefont {P.~J.}\ \bibnamefont
  {{Marshall}}}, \bibinfo {author} {\bibfnamefont {M.~W.}\ \bibnamefont
  {{Auger}}}, \bibinfo {author} {\bibfnamefont {T.}~\bibnamefont {{Treu}}},
  \bibinfo {author} {\bibfnamefont {D.~C.}\ \bibnamefont {{Koo}}}, \bibinfo
  {author} {\bibfnamefont {A.~S.}\ \bibnamefont {{Bolton}}}, \bibinfo {author}
  {\bibfnamefont {B.~P.}\ \bibnamefont {{Holden}}}, \ and\ \bibinfo {author}
  {\bibfnamefont {L.~V.~E.}\ \bibnamefont {{Koopmans}}},\ }\href {\doibase
  10.1111/j.1365-2966.2011.18706.x} {\bibfield  {journal} {\bibinfo  {journal}
  {\mnras}\ }\textbf {\bibinfo {volume} {417}},\ \bibinfo {pages} {1621}
  (\bibinfo {year} {2011})},\ \Eprint {http://arxiv.org/abs/1101.1622}
  {arXiv:1101.1622 [astro-ph.CO]} \BibitemShut {NoStop}%
\bibitem [{\citenamefont {{More}}\ \emph {et~al.}(2012)\citenamefont {{More}},
  \citenamefont {{Cabanac}}, \citenamefont {{More}}, \citenamefont {{Alard}},
  \citenamefont {{Limousin}}, \citenamefont {{Kneib}}, \citenamefont
  {{Gavazzi}},\ and\ \citenamefont {{Motta}}}]{more11}%
  \BibitemOpen
  \bibfield  {author} {\bibinfo {author} {\bibfnamefont {A.}~\bibnamefont
  {{More}}}, \bibinfo {author} {\bibfnamefont {R.}~\bibnamefont {{Cabanac}}},
  \bibinfo {author} {\bibfnamefont {S.}~\bibnamefont {{More}}}, \bibinfo
  {author} {\bibfnamefont {C.}~\bibnamefont {{Alard}}}, \bibinfo {author}
  {\bibfnamefont {M.}~\bibnamefont {{Limousin}}}, \bibinfo {author}
  {\bibfnamefont {J.-P.}\ \bibnamefont {{Kneib}}}, \bibinfo {author}
  {\bibfnamefont {R.}~\bibnamefont {{Gavazzi}}}, \ and\ \bibinfo {author}
  {\bibfnamefont {V.}~\bibnamefont {{Motta}}},\ }\href {\doibase
  10.1088/0004-637X/749/1/38} {\bibfield  {journal} {\bibinfo  {journal}
  {\apj}\ }\textbf {\bibinfo {volume} {749}},\ \bibinfo {eid} {38} (\bibinfo
  {year} {2012})},\ \Eprint {http://arxiv.org/abs/1109.1821} {arXiv:1109.1821
  [astro-ph.CO]} \BibitemShut {NoStop}%
\bibitem [{\citenamefont {{Newman}}\ \emph {et~al.}(2009)\citenamefont
  {{Newman}}, \citenamefont {{Treu}}, \citenamefont {{Ellis}}, \citenamefont
  {{Sand}}, \citenamefont {{Richard}}, \citenamefont {{Marshall}},
  \citenamefont {{Capak}},\ and\ \citenamefont {{Miyazaki}}}]{newman09}%
  \BibitemOpen
  \bibfield  {author} {\bibinfo {author} {\bibfnamefont {A.~B.}\ \bibnamefont
  {{Newman}}}, \bibinfo {author} {\bibfnamefont {T.}~\bibnamefont {{Treu}}},
  \bibinfo {author} {\bibfnamefont {R.~S.}\ \bibnamefont {{Ellis}}}, \bibinfo
  {author} {\bibfnamefont {D.~J.}\ \bibnamefont {{Sand}}}, \bibinfo {author}
  {\bibfnamefont {J.}~\bibnamefont {{Richard}}}, \bibinfo {author}
  {\bibfnamefont {P.~J.}\ \bibnamefont {{Marshall}}}, \bibinfo {author}
  {\bibfnamefont {P.}~\bibnamefont {{Capak}}}, \ and\ \bibinfo {author}
  {\bibfnamefont {S.}~\bibnamefont {{Miyazaki}}},\ }\href {\doibase
  10.1088/0004-637X/706/2/1078} {\bibfield  {journal} {\bibinfo  {journal}
  {\apj}\ }\textbf {\bibinfo {volume} {706}},\ \bibinfo {pages} {1078}
  (\bibinfo {year} {2009})},\ \Eprint {http://arxiv.org/abs/0909.3527}
  {arXiv:0909.3527 [astro-ph.CO]} \BibitemShut {NoStop}%
\bibitem [{\citenamefont {{Limousin}}\ \emph {et~al.}(2007)\citenamefont
  {{Limousin}}, \citenamefont {{Richard}}, \citenamefont {{Jullo}},
  \citenamefont {{Kneib}}, \citenamefont {{Fort}}, \citenamefont {{Soucail}},
  \citenamefont {{El{\'{\i}}asd{\'o}ttir}}, \citenamefont {{Natarajan}},
  \citenamefont {{Ellis}}, \citenamefont {{Smail}}, \citenamefont {{Czoske}},
  \citenamefont {{Smith}}, \citenamefont {{Hudelot}}, \citenamefont
  {{Bardeau}}, \citenamefont {{Ebeling}}, \citenamefont {{Egami}},\ and\
  \citenamefont {{Knudsen}}}]{limousin07}%
  \BibitemOpen
  \bibfield  {author} {\bibinfo {author} {\bibfnamefont {M.}~\bibnamefont
  {{Limousin}}}, \bibinfo {author} {\bibfnamefont {J.}~\bibnamefont
  {{Richard}}}, \bibinfo {author} {\bibfnamefont {E.}~\bibnamefont {{Jullo}}},
  \bibinfo {author} {\bibfnamefont {J.-P.}\ \bibnamefont {{Kneib}}}, \bibinfo
  {author} {\bibfnamefont {B.}~\bibnamefont {{Fort}}}, \bibinfo {author}
  {\bibfnamefont {G.}~\bibnamefont {{Soucail}}}, \bibinfo {author}
  {\bibfnamefont {{\'A}.}~\bibnamefont {{El{\'{\i}}asd{\'o}ttir}}}, \bibinfo
  {author} {\bibfnamefont {P.}~\bibnamefont {{Natarajan}}}, \bibinfo {author}
  {\bibfnamefont {R.~S.}\ \bibnamefont {{Ellis}}}, \bibinfo {author}
  {\bibfnamefont {I.}~\bibnamefont {{Smail}}}, \bibinfo {author} {\bibfnamefont
  {O.}~\bibnamefont {{Czoske}}}, \bibinfo {author} {\bibfnamefont {G.~P.}\
  \bibnamefont {{Smith}}}, \bibinfo {author} {\bibfnamefont {P.}~\bibnamefont
  {{Hudelot}}}, \bibinfo {author} {\bibfnamefont {S.}~\bibnamefont
  {{Bardeau}}}, \bibinfo {author} {\bibfnamefont {H.}~\bibnamefont
  {{Ebeling}}}, \bibinfo {author} {\bibfnamefont {E.}~\bibnamefont {{Egami}}},
  \ and\ \bibinfo {author} {\bibfnamefont {K.~K.}\ \bibnamefont {{Knudsen}}},\
  }\href {\doibase 10.1086/521293} {\bibfield  {journal} {\bibinfo  {journal}
  {\apj}\ }\textbf {\bibinfo {volume} {668}},\ \bibinfo {pages} {643} (\bibinfo
  {year} {2007})},\ \Eprint {http://arxiv.org/abs/arXiv:astro-ph/0612165}
  {arXiv:astro-ph/0612165} \BibitemShut {NoStop}%
\bibitem [{\citenamefont {{Bernal}}\ \emph {et~al.}(2011)\citenamefont
  {{Bernal}}, \citenamefont {{Capozziello}}, \citenamefont {{Hidalgo}},\ and\
  \citenamefont {{Mendoza}}}]{bernal11}%
  \BibitemOpen
  \bibfield  {author} {\bibinfo {author} {\bibfnamefont {T.}~\bibnamefont
  {{Bernal}}}, \bibinfo {author} {\bibfnamefont {S.}~\bibnamefont
  {{Capozziello}}}, \bibinfo {author} {\bibfnamefont {J.~C.}\ \bibnamefont
  {{Hidalgo}}}, \ and\ \bibinfo {author} {\bibfnamefont {S.}~\bibnamefont
  {{Mendoza}}},\ }\href {\doibase 10.1140/epjc/s10052-011-1794-z} {\bibfield
  {journal} {\bibinfo  {journal} {Eur. Phys. J. C}\ }\textbf {\bibinfo {volume}
  {71}},\ \bibinfo {pages} {1794} (\bibinfo {year} {2011})},\ \Eprint
  {http://arxiv.org/abs/1108.5588} {arXiv:1108.5588 [astro-ph.CO]} \BibitemShut
  {NoStop}%
\end{thebibliography}%

\end{document}